\documentclass[conference]{IEEEtran}

\ifCLASSINFOpdf
\else
\fi
\usepackage{url}

\usepackage[]{fancyhdr} %
\newcommand{\changefont}{\fontsize{9}{9}\selectfont}
\IEEEoverridecommandlockouts
\usepackage{cite}
\usepackage{amsmath,amssymb,amsfonts}
\usepackage{algorithmic}
\usepackage{graphicx}
\usepackage{textcomp}
\usepackage{xcolor}
\usepackage{tabularx}
\usepackage{tabulary}
\makeatletter
\def\TY@box@v#1{%
      $\vcenter \@startpbox{\csname TY@F\the\TY@count\endcsname}%
              #1\arraybackslash\tyformat
                              \insert@column\@endpbox$}
\makeatother
\def\BibTeX{{\rm B\kern-.05em{\sc i\kern-.025em b}\kern-.08em
    T\kern-.1667em\lower.7ex\hbox{E}\kern-.125emX}}

\usepackage[newcommands]{ragged2e}
\newtheorem{remark}{Remark}
\begin{document}

%
\title{Consensus-based Frequency and Voltage Regulation for Fully Inverter-based Islanded Microgrids}

\author{\IEEEauthorblockN{Y. Cheng\textsuperscript{1,2}, Tao Liu\textsuperscript{1,2}}
\IEEEauthorblockA{1. Department of Electrical \\and Electronic Engineering\\The University of Hong Kong\\
Hong Kong SAR, China\\ 2. Shenzhen Institute of\\Research and Innovation\\The University of Hong Kong}
\and
\IEEEauthorblockN{David J. Hill\textsuperscript{3,4}}
\IEEEauthorblockA{3. Department of Electrical \\and Electronic Engineering\\The University of Hong Kong\\
Hong Kong SAR, China\\4. School of Electrical Engineering\\ and Telecommunications\\University of New South Wales\\Sydney, Australia}
\and
\IEEEauthorblockN{Xue Lyu}
\IEEEauthorblockA{Department of Electrical \\and Computer Engineering\\University of Wisconsin–Madison\\
Madison, USA\\
} \thanks{This work was supported by the Research Grants Council of the Hong Kong Special Administrative Region under the General Research Fund through Project No. 17209219, and the National Natural Science Foundation of China
through project No. 62173287.}}


%





\maketitle
\thispagestyle{fancy}
\pagestyle{fancy}


\begin{abstract}
This paper proposes a new distributed consensus-based control method for voltage and frequency control of fully inverter-based islanded microgrids (MGs). The proposed method includes the active power sharing in voltage control to improve the reactive power sharing accuracy and thus generalizes some existing secondary frequency and voltage control methods. Firstly, frequency is regulated by distributed secondary frequency control. Secondly, voltage is regulated by distributed average voltage control and decentralized individual voltage control. It offers a tunable trade-off between voltage regulation, active, and reactive power sharing accuracy. Therefore, it avoids the abuse of sacrificing reactive power sharing accuracy for exact voltage regulation which is a common issue of existing methods. The proposed method is implemented in a distributed way that does not require a prior knowledge of the MG network structure and loads and hence can ensure scalability. Simulation results shows that the proposed controller achieves different compromise between the above three targets under different modes of operation.
\end{abstract}


\section{Introduction}
Driven by the climate change and global warming, power systems are undergoing a significant transition. The core of this transition is to gradually replace fossil fuel based synchronous generators with renewable energy source-powered generators that are usually connected to the grid via converters, i.e., the converter interfaced generators (CIGs). The concept of MG plays an important role in connecting these CIGs, and thus attracts increasing attention recently. Usually, a MG consists of distributed generators such as wind turbines, solar panels, and microturbines \cite{A} that are connected to the MG via inverters. Therefore, the control of a fully inverter-based islanded MG is of great importance.

Conventional $P-\omega$, $Q-V$ droop control method is the most popular primary control in fully inverter-based MGs since it is simple to implement and does not require communication between CIGs. But its drawbacks cannot be neglected. Reference \cite{C} summarizes these drawbacks. The most critical one is that it cannot accurately share reactive power among CIGs due to the mismatch in output impedance, which may cause circulating current among CIGs \cite{C,D,E,F,G,J}. Some new control methods have been developed to address the issue, e.g., the $Q-\dot{V}$ droop control in \cite{F} and the virtual impedance method in \cite{E,G}. These methods require {\it a prior} knowledge of the MG network. In addition, $P-V$, $Q-\omega$ droop control is also used in the literature due to the high R/X ratio of MGs \cite{H}.

On the other hand, droop control results in voltage and frequency deviations and thus secondary control methods are needed to restore their nominal values, e.g., \cite{I1,I2,L}. However, many of the existing secondary voltage control methods aim at exact voltage regulation, i.e., restoring all CIG output voltage magnitude to a common reference value. \cite{I1,L}. This may further deteriorate the reactive power sharing accuracy due to the fundamental conflict between voltage regulation and reactive power sharing \cite{J}. To solve this issue, a tunable controller is proposed to achieve a balance between the accuracy in reactive power sharing and the performance in voltage regulation in \cite{J}. Nevertheless, apart from reactive power, active power changes may also influence bus voltages in a MG due to its high R/X ratio and thus can be an alternative candidate to contribute to voltage regulation. To get active power in voltage regulation, trade-off between the accuracy in active power sharing and reactive power sharing, and the performance in voltage regulation should be considered simultaneously in the controller design of MGs to relax the burden of reactive power sharing accuracy.

Inspired by \cite{J}, this paper proposes a new tunable control method by including active power sharing in voltage regulation to achieve compromise among the three aforementioned control targets. We use reactive power control to regulate the frequency, and both active and reactive power control to regulate voltages. For the frequency regulation, we use the $Q-\omega$ droop control for the primary frequency control and a distributed secondary frequency control to restore the frequency to nominal value. For the voltage control, we offer an average voltage regulation and a triggered individual voltage regulation. The former is aimed to regulate average CIG voltage magnitude by controlling their active power output according to the active power sharing ratio. Only regulating average voltage may not be sufficient as individual CIG voltage magnitude may still be out of acceptable range. Therefore, we offer the triggered individual voltage regulation control method for each CIG to trigger when some preset conditions are satisfied, e.g., its output voltage is out of acceptable range. In individual voltage regulation, the active and reactive power output of that CIG is controlled without the consideration of power sharing ratio to regulate its output voltage magnitude.

The novelties of the proposed control method are threefold. Firstly, it can be implemented in distributed way and thus is scalable without requiring {\it  a prior} knowledge of the MG's network structure and loads. Secondly, the power sharing errors stemming from output impedance mismatch in conventional voltage droop control is solved by the proposed method where the corresponding power variable setpoints of CIGs, i.e., active power in this case, are determined via consensus-based control to achieve accurate power sharing. The CIGs' output voltage magnitudes are used to control to the desire power injection. This is different from voltage droop control where the output voltage magnitude follows the changes of power output. Effect of physical quantities of the MG, i.e., output impedance, in power sharing in voltage droop control is eliminated as the consensus is reached on communication level. Thirdly, it generalizes the methods proposed  in \cite{I1,I2,L,J} to achieve a trade-off between the aforementioned three control targets. In particular, it extends the ideas in \cite{J} to include the accuracy of active power sharing in the trade-off to prevent abusing the reactive power sharing accuracy.

The rest of the paper is organized as follows. Section II introduces the system model. In section III, the proposed control method is introduced and its performance is demonstrated in Section IV. Conclusions are given in Section V.

\section{Model Description}
We consider a MG with $n$ buses and $m$ distribution lines with $\mathcal{N}=\{1,\dots,n\}$, $\mathcal{M}=\{1,\dots,m\}$ as the corresponding index sets. We divide all buses in the MG into three types: a CIG bus, a load bus, and an intermediate bus that refers to a bus that has no CIGs and loads. We denote the index set of CIG buses as $\mathcal{C}=\{1, 2,\dots,c\}$, and load bus as $\mathcal{L}=\{c+1,\dots,l\}$, where $c$ and $l$ are the number of CIGs and loads in the MG respectively. For simplicity, we assume $\mathcal{C}\cap\mathcal{L}=\varnothing $ and the MG has at least one CIG, one  line, and one load.

Here we only consider a three-phase balanced MG. We apply Park transformation to transform three-phase balanced signals into the direct and quadrature ($d-q$) axes components. We denote the CIG at bus $i$, $i\in \mathcal{C}$, as $CIG_i$ and its model is built under the $d_i-q_i$ frame at its local frequency $\omega_i$. For each  line $j$, $j\in \mathcal{M}$ and each load bus $h$, $h\in\mathcal{L}$, their models are built under the system $D-Q$ frame at the nominal frequency $\omega_s$. To form the whole network, we change a CIG variable $\boldsymbol{x_i^{dq}}$ under the $d_i-q_i$ frame to that $\boldsymbol{X_i^{DQ}}$ under the $D-Q$ frame by a transformation $T_i$, i.e., $\boldsymbol{X_i^{DQ}}=T_i\boldsymbol{x_i^{dq}}$ and $\boldsymbol{x_i^{dq}}=T_i^{-1}\boldsymbol{X_i^{DQ}}$ with $\boldsymbol{x_i^{dq}}=(x_i^d,x_i^q)^\top$, $\boldsymbol{X_i^{DQ}}=(X_i^D,X_i^Q)^\top$, \cite{K,E}. The transformation matrix $T_i$ is defined as:
\begin{equation}
T_i=\begin{pmatrix}
\cos\delta_i & -\sin\delta_i \\ 
\sin\delta_i & \cos\delta_i
\end{pmatrix}\\
\end{equation}
with $\dot{\delta}_i=\omega_i-\omega_s$.

The superscripts $\boldsymbol{dq}$ and $\boldsymbol{DQ}$ denote that the rated variables are defined under the $d_i-q_i$ and $D-Q$ frame respectively. The whole MG model consists of differential algebraic equations. The differential equations are used to represent dynamics of CIGs, lines, and loads are described below. 
\begin{figure}[htbp]
\centerline{\includegraphics[scale=0.80]{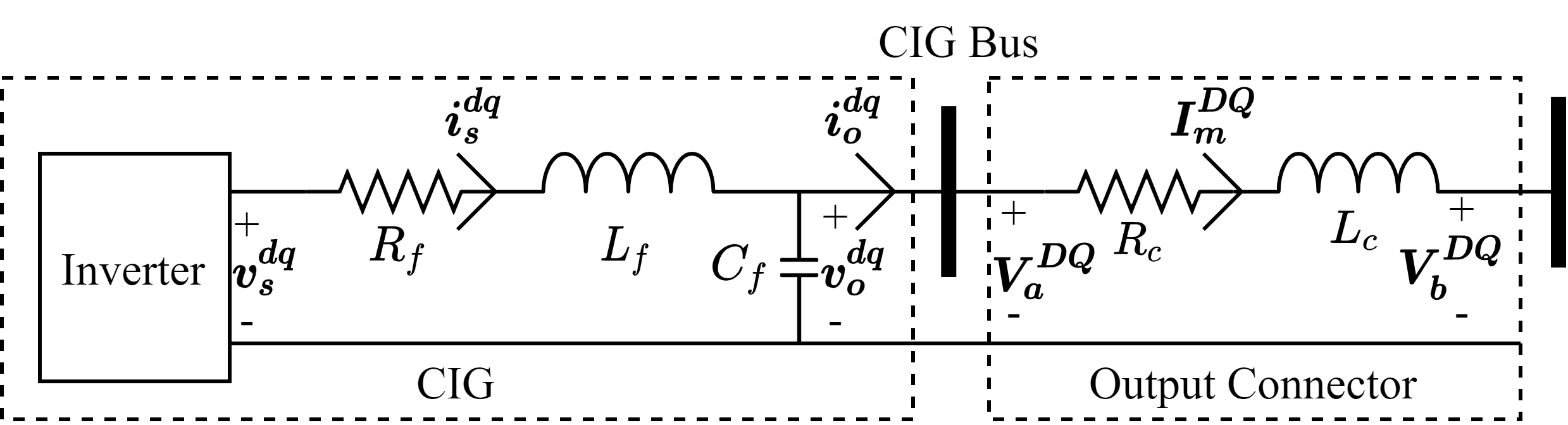}}
\caption{Configuration of a CIG and its output connector.}
\label{fig1}
\end{figure}

\subsection{CIG Model}

Here, we adopt the configuration of CIGs as that in \cite{I1,I2,L,K}. For simplicity, we neglect the DC-side dynamics for simplicity, and adopt the assumption therein that each CIG connects to the MG via an output connector which is modelled as a distribution line with resistance $R_c$ and inductance $L_c$. The configuration of the CIG and its ouput connector is shown in Fig. \ref{fig1}, where the switching voltage $\boldsymbol{v_{s_i}^{dq}}$ and frequency $\omega_i$ are considered as input variables, which will be designed later. The mathematical model of $CIG_i$ with the switching current $\boldsymbol{i_{s_i}^{dq}}$ and $\boldsymbol{v_{o_i}^{dq}}$ as its states are given as follows \cite{I1,I2,L,K}:
\begin{align}
\boldsymbol{\dot{i}_{s_i}^{dq}}&=L_{f_i}^{-1}(-R_{f_i}\boldsymbol{i_{s_i}^{dq}+\boldsymbol{v_{s_i}^{dq}}-\boldsymbol{v_{o_i}^{dq}}})+\omega_iK\boldsymbol{i_{s_i}^{dq}}\\
\boldsymbol{\dot{v}_{o_i}^{dq}}&=C_{f_i}^{-1}(\boldsymbol{i_{s_i}^{dq}}-\boldsymbol{i_{o_i}^{dq}})+\omega_iK\boldsymbol{v_{o_i}^{dq}}
\end{align}
where $K=\begin{bmatrix}
0 & 1\\ 
-1 & 0
\end{bmatrix}$; $R_{f_i}$, $L_{f_i}$, $C_{f_i}$ are the output filter's resistance, inductance, and capacitance of $CIG_i$, respectively.

\subsection{Distribution Line Model}

For a distribution line $m$, $m\in \mathcal{M}$, that connects bus $a$ and $b$ with $a,b\in \mathcal{N}$, we model it as a $RL$ circuit with a series resistance $R_m$ and inductance $L_m$. The dynamics of the current \boldmath $I_{m}^{DQ}$ \unboldmath going through the line is \cite{K}:
\begin{align}
\boldsymbol{\dot{I}_{m}^{DQ}}&=L_m^{-1}(-R_m\boldsymbol{I_{m}^{DQ}}+\boldsymbol{V_{a}^{DQ}}-\boldsymbol{V_{b}^{DQ}})+\omega_sK\boldsymbol{I_{m}^{DQ}}
\end{align}
where \boldmath$V_{a}^{DQ}$ and $V_{b}^{DQ}$ \unboldmath are the bus voltages at bus $a$ and $b$, respectively.

\subsection{Load Model}

We assume that the load at bus $h$, $h\in\mathcal{L}$, is a series $RL$ load with a resistance $R_{load_h}$ and inductance $L_{load_h}$. The dynamics of the current, $\boldsymbol{\dot{I}_h^{DQ}}$, going through the load is \cite{K}:
\begin{align}
\boldsymbol{\dot{I}_h^{DQ}}&=L_{load_h}^{-1}(-R_{load_h}\boldsymbol{I_h^{DQ}}+\boldsymbol{V_h^{DQ}})+\omega_sK\boldsymbol{I_h^{DQ}}.
\end{align}
Note that the voltage at the load bus, $\boldsymbol{V_h^{DQ}}$, can be calculated by solving the algebraic equation established by Kirchhoff's Current Law (KCL) at the load bus $h$.

Different components are connected together by applying KCL at every buses to establish the algebraic equations of the whole network.
\begin{figure*}[t]
\centering
\includegraphics[width=\linewidth]{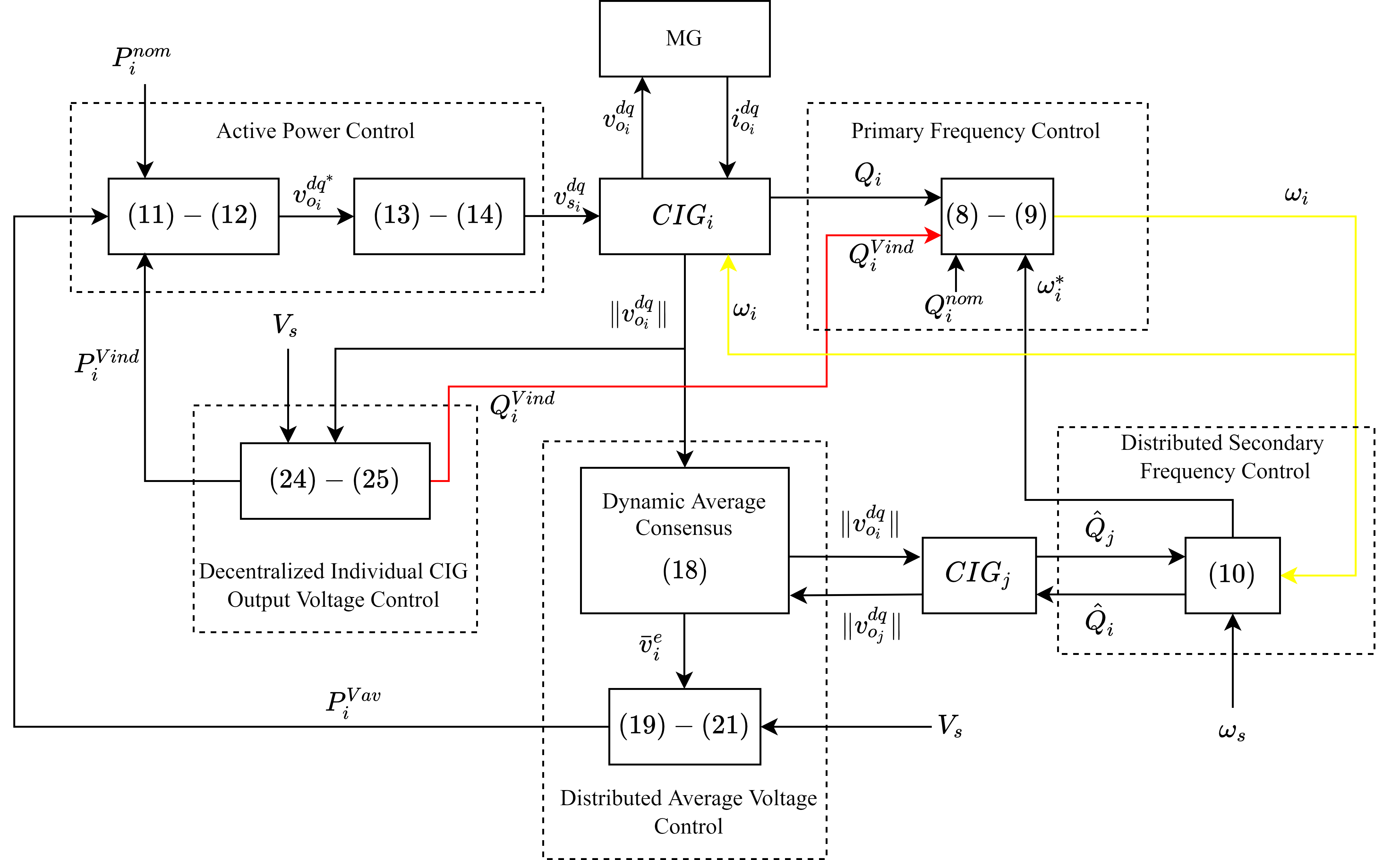}
\caption{Control block diagram of proposed controller}
\end{figure*}
\section{Proposed Control Method}
The proposed control method is aimed to regulate the frequency, voltage, and power sharing accuracy by controlling $CIG_i$ output frequency $\omega_i$, its reactive power output $Q_i$, and active power output $P_i$. The setpoint $P_i^*$ of $P_i$ is the sum of three variables, i.e., the nominal active power output $P_i^{nom}$, the active power $P_i^{Vav}$ to regulate average CIG output magnitude $\|\bar{v}_o\|$, and the active power $P_i^{Vind}$ to regulate individual $CIG_i$ output voltage magnitude $\|\boldsymbol{v_{o_i}^{dq}}\|$.

For frequency regulation, $Q-\omega$ droop control and its associated distributed secondary frequency control are used. For voltage regulation, it offers the average voltage regulation and individual voltage regulation. In average voltage regulation, distributed consensus algorithm is implemented to give an estimation of average $CIG_i$ voltage magnitude, $\|\bar{v}_o\|$, to each $CIG_i$. Then, $CIG_i$ calculates $P_i^{Vav}$ based on predefined active power sharing ratio to regulate $\|\bar{v}_o\|$. As accurate power sharing conflict with exact voltage regulation, only exact average voltage regulation can be achieved with accurate power sharing. If individual voltage regulation is required, i.e., even when average voltage magnitude equals to nominal value, we still require the output voltage magnitude of a CIG to get closer to nominal value or even be equal to nominal value, then the power output of that CIG needs to change without considering power sharing ratio. Therefore, in individual voltage regulation, $CIG_i$ calculates $P_i^{Vind}$ and $Q_i^{Vind}$ regardless of power sharing ratio to control $P_i$ and $Q_i$ to regulate $\|\boldsymbol{v_{o_i}^{dq}}\|$. Summing $P_i^{nom}$, $P_i^{Vav}$, and $P_i^{Vind}$, the setpoint $P_i^*$ of $P_i$ is obtained. Then a cascaded inner control loop is implemented to calculate the control input to $CIG_i$, i.e., the switching voltage, to inject the desired active power to MG. Fig 2. shows the control block diagram of the proposed control method. We assume $P_i^{nom}$ and $Q_i^{nom}$ are selected based on the predefined power sharing ratio.

The instantaneous active power output $P_i$, reactive power output $Q_i$, and output voltage magnitude $\|\boldsymbol{v_{o_i}^{dq}}\|$ of $CIG_i$ are calculated as follows:
\begin{align}
P_i=v_{o_i}^di_{o_i}^d+v_{o_i}^qi_{o_i}^q&, Q_i=v_{o_i}^qi_{o_i}^d-v_{o_i}^di_{o_i}^q,\label{eq3}\\
 \|\boldsymbol{v_{o_i}^{dq}}\|&=\sqrt{{(v_{o_i}^d)}^2+{(v_{o_i}^q)}^2}.\label{eq4}
\end{align}
\subsection{Primary Frequency Control: $Q-\omega$ Droop Control}

The output frequency $\omega_i$ of $CIG_i$ is controlled as follows \cite{Q}: 
\begin{align}
Q_i^{base}&=Q_i^{nom}+Q_i^{Vind}\\
\omega_i&=\omega_i^*+m_i(Q_i-Q_i^{base})
\end{align}
where $Q_i^{base}$ is the reactive power output base value and is the sum of the nominal reactive power base value $Q_i^{nom}$ and reactive power $Q_i^{Vind}$ to regulate individual $CIG_i$ output voltage magnitude. $\omega_i^*$ is the frequency setpoint to be determined in section \ref{sectionB}; $m_i$ is the frequency droop coefficient of $CIG_i$. $Q-\omega$ droop control results in frequency deviation and thus distributed secondary frequency control is used to control it back to $\omega_s$ by controlling $\omega_i^*$.

\subsection{Distributed Secondary Frequency Control} \label{sectionB}
We assume that there exists a bidirectional connected communication network between CIGs. We denote its network topology as an undirected graph $\mathcal{G}$. We use $A\in\mathbb{R}^{c\times c}$ to denote the adjacency matrix of $\mathcal{G}$ with all edge weights equal to one and $a_{ij}$ denotes the $(i,j)$th entry of $A$. We assume the communication between CIGs is periodic with a period $t_1$, i.e., at every $t_1$ seconds, each $CIG_i$ sends its current output frequency $\hat{\omega}_{i,k_1}=\omega_i(k_1t_1)$, reactive power sharing signals $\hat{Q}_{i,k_1}=m_i(Q_i(k_1t_1)-Q_i^{base}(k_1t_1))$ with $k_1=0,1,2,\dots$ to its communication neighbours. As $Q_i^{base}$ is included in the droop control, the distributed secondary frequency control in \cite{H} is modified accordingly and the local setpoint $\omega_i^*$ of $CIG_i$ is calculated as follows:
\begin{align}\label{eq1}
\dot{\omega}_i^*=&-c_f(b_i(\omega_i-\omega_s)+\sum_{j\in \mathcal{N}_i}a_{ij}(\omega_i-\hat{\omega}_{j,k_1}\nonumber\\
&-m_iQ_i+\hat{Q}_{j,k_1}))
\end{align}
where $\mathcal{N}_i$ is the index set of communication neighbours of $CIG_i$; $c_f$ and $b_i$ are the the coupling gain and pinning gain, respectively. We assume $b_i\geq 0$ is non-zero for only one selected CIG. The distributed secondary frequency control was originally proposed in \cite{I2} for $P-\omega$ droop. Here, we modified it for the $Q-\omega$ droop \cite{H,I2} in \eqref{eq1}. The term $a_{ij}(\omega_i-\hat{\omega}_{j,k_1}))+b_i(\omega_i-\omega_s)$ is the tracking error of frequency while $a_{ij}(-m_i(Q_i-Q_i^{base})+\hat{Q}_{j,k_1})$ is the tracking error of reactive power sharing. Their convergence speed can be adjusted independently by assigning two different gains but here we use a common coupling gain $c_f$ for simplicity. The term $b_i(\omega_i-\omega_s)$ is to ensure that one CIG has access to $\omega_s$ and its local frequency is tracked with $\omega_s$ with the tracking rate determined by $b_i$. 

\subsection{Active Power Control}

The aim of active power control is to regulate the active power output of $CIG_i$ $P_i$ to its setpoint $P_i^*$ which consists of three components, i.e., $P_i^{nom}$, $P_i^{Vav}$, and $P_i^{Vind}$. In the rest of the paper, we use  $K_p$ and $K_i$ to denote the proportional gain and integral gain of a PI controller respectively with different superscripts representing different PI controllers. The setpoint $P_i^*$, $v_{o_i}^d$ setpoint $v_{o_i}^{d^*}$ are calculated as follows:
\begin{align}
P_i^*&=P_i^{nom}+P_i^{Vav}+P_i^{Vind}\\
v_{o_i}^{d^*}&=K_p^{vd}(v_{o_i}^{d^{set}}-v_{o_i}^d)+K_i^{vd}\int_0^t (v_{o_i}^{d^{set}}-v_{o_i}^d)dt.
\end{align}
The control variable $v_{o_i}^{d^{set}}$ is defined as $v_{o_i}^{d^{set}}=sgn(i_{o_i}^d)\frac{P_i^*}{(\left|i_{o_i}^d\right|+\varepsilon)}$, where $\varepsilon$ is a small positive constant to avoid $v_{o_i}^{d^{set}}$ going to infinity; $sgn(i_{o_i}^d)=1$ for $i_{o_i}^d\geq0$ and $sgn(i_{o_i}^d)=-1$ otherwise. $v_{o_i}^q$ setpoint $v_{o_i}^{q^*}$ is set to zero. The $\boldsymbol{v_{o_i}^{dq}}$ setpoint $\boldsymbol{v_{o_i}^{dq^*}}$ are then used to calculate the control input, i.e., switching voltage $\boldsymbol{v_{s_i}^{dq}}$, of $CIG_i$ as follows to regulate $\boldsymbol{v_{o_i}^{dq}}$ to $\boldsymbol{v_{o_i}^{dq^*}}$ \cite{I1,L,K}:
\begin{align}
\boldsymbol{i_{s_i}^{dq^*}}=&F\boldsymbol{i_{o_i}^{dq}}-C_f\omega K\boldsymbol{v_{o_i}^{dq}}+K_p^v(\boldsymbol{v_{o_i}^{dq^*}-\boldsymbol{v_{o_i}^{dq}}})\nonumber\\
&+K_i^v\int_0^t(\boldsymbol{v_{o_i}^{dq^*}-\boldsymbol{v_{o_i}^{dq}}})dt\\
\boldsymbol{v_{s_i}^{dq}}=&-L_f\omega K\boldsymbol{i_{s_i}^{dq}}+K_p^i(\boldsymbol{i_{s_i}^{dq^*}-\boldsymbol{i_{s_i}^{dq}}})\nonumber\\
&+K_i^i\int_0^t(\boldsymbol{i_{s_i}^{dq^*}-\boldsymbol{i_{s_i}^{dq}}})dt
\end{align}
where $F$ is the current feed forward gain.
\subsection{Distributed Average Voltage Control}
We regulate $\|\bar{v}_o\|$ by controlling $P_i^{Vav}$ based on predefined active power sharing ratio. The following shows the correlation between $P_i$ and $\|\bar{v}_o\|$ in steady state.

The active power flow equation of $CIG_i$ in bus $i$ is as follows:
\begin{align}
P_i=\|V_i\|\|V_j\|(-G\cos(\theta_i-\theta_j)-B\sin(\theta_i-\theta_j))+G\|V_i\|^2
\end{align}
where $\|V_i\|$ and $\theta_i$ are the voltage magnitude and phase angle of bus $i$, respectively; $G$ and $B$ are the conductance and susceptance of output connector.

Taking the partial derivative of $P_i$ with respect to $\|V_i\|$, we have the following:
\begin{align}
\frac{\partial P_i}{\partial \|V_i\|}=\frac{P_i+G\|V_i\|^2}{\|V_i\|}.\label{eq5}
\end{align}
As $P_i\geq0$, $G>0$ and $\|V_i\|^2\geq0$, we can conclude that $P_i$ has a positive correlation with $\|V_i\|$, i.e., $\|\boldsymbol{v_{o_i}^{dq}}\|$.

The following shows the definition of $\|\bar{v}_o\|$:
\begin{align}
\|\bar{v}_o\|=\frac{1}{c}\sum_{i=1}^c\|\boldsymbol{v_{o_i}^{dq}}\|.
\end{align}
It can be easily conclude that $P_i$ has a positive correlation with $\|\boldsymbol{v_{o_i}^{dq}}\|$ and thus $\|\bar{v}_o\|$.

If every $CIG_i$ knows $\|\bar{v}_o\|$ and changes their $P_i$ based on $\|\bar{v}_o\|$ and predefined active power sharing ratio to regulate $\|\bar{v}_o\|$, then $\|\bar{v}_o\|$ can be driven to nominal CIG output voltage magnitude, $V_s$, while maintaining accurate active power sharing among CIGs. Thus, dynamic average consensus algorithm is implemented to give an estimation $\|\bar{v}_o\|$ in the MG to each $CIG_i$. At every sampling time $k_2t_2$ with $k_2=0,1,2,\dots$ and $t_2>t_1$, $CIG_i$ sets the input $u_{i,k_2}$ to the consensus algorithm as $\|\boldsymbol{v_{o_i}^{dq}}(k_2t_2)\|$ and fixes it during that sampling time interval. $CIG_i$ communicates every $t_1$ seconds and the state variable $r_i$ and the agreement state $x_i$ of $CIG_i$ are calculated as follows \cite{N}:
\begin{align}
r_{i,k_1+1}&=(1+\rho^2)r_{i,k_1}-\rho^2r_{i,k_1-1}\nonumber\\
&+k_I\sum_{j=1}^Na_{ij}(x_{i,k_1}-x_{j,k_1})
\end{align}
where $x_{i,k_1}=u_{i,k_2}-r_{i,k_1}$, $\rho$ and $k_I$ are parameters. $r_{i,0}$ is initialized as zero to satisfy to requirement of the average of initial integrator states equal to zero \cite{N}. At sampling time $k_2t_2$, $x_{i,k_1}$ is considered as the estimation of average inverter bus voltage magnitude $\bar{v}_{i,k_2}$ for $CIG_i$ first and then $u_{i,k_2}$ is updated with $\|\boldsymbol{v_{o_i}^{dq}}(k_2t_2)\|$.

The active power contribution of $CIG_i$ in regulating $\|\bar{v}_o\|$, $P_i^{Vav}$, which is a component of $P_i^*$, is controlled by the discrete PI controller with the compensator formula $K_p^{A}+K_i^{A}\frac{t_2}{z-1}$ and sample time $t_2$ as follows:
\begin{align}
e_{i,k_2}&=V_n-\bar{v}_{i,k_2}^e\\
d_{i,k_2}&=d_{i,k_2-1}+K_p^{A}(e_{i,k_2}-e_{i,k_2-1})+K_i^{A}(t_2\cdot e_{i,k_2-1})\\
P_i^{Vav}&=n_id_{i,k_2}
\end{align}
where $n_i$ denote the $P_i^{Vav}$ sharing gain of $CIG_i$. As long as every CIG receives the same $\bar{v}_{i,k_2}^e$ signal, i.e., consensus is reached, the responsibility of regulating $\|\bar{v}_o\|$ via active power can be shared accurately based on $n_i$. The number of iterations needed for the average consensus error to be smaller than certain threshold depends on the topology of the communication network. Thus, $t_2$ can be determined based on $t_1$ and the acceptable $P_i^{Vav}$ sharing error. \cite{N} provides detailed analysis of the convergence rate of the above algorithm and the rules of choosing optimal parameters.

\subsection{Decentralized Individual CIG Output Voltage Control}
Decentralized individual CIG output voltage control is a triggered control function for individual CIG and is triggered under some preset conditions, e.g., the $\|\boldsymbol{v_{o_i}^{dq}}\|$ is out of acceptable range. Two variables, $P_i^{Vind}$ and $Q_i^{Vind}$, are controlled to vary $P_i$ and $Q_i$ to regulate $\|\boldsymbol{v_{o_i}^{dq}}\|$. As mentioned before, the accurate power sharing and individual $\|\boldsymbol{v_{o_i}^{dq}}\|$ voltage regulation conflict with each other. Therefore, the predefined power sharing ratio is not considered in individual $\|\boldsymbol{v_{o_i}^{dq}}\|$ voltage regulation, i.e., $P_i^{Vind}$ and $Q_i^{Vind}$ are calculated without taking the power sharing ratio into account. When it is triggered, $\|\boldsymbol{v_{o_i}^{dq}}\|$ is regulated and the power sharing accuracy is not guaranteed. When it is not triggered, $P_i^{Vind}$ and $Q_i^{Vind}$ are set to zero and power sharing accuracy is preserved.

The positive correlation between $P_i$ and $\|\boldsymbol{v_{o_i}^{dq}}\|$ has been shown in \eqref{eq5} and the correlation between $Q_i$ and $\|\boldsymbol{v_{o_i}^{dq}}\|$ in steady state is showed as the following.

The reactive power flow equation of $CIG_i$ in bus $i$ is as follows:
\begin{align}
Q_i=&\|V_i\|\|V_j\|(-G\sin(\theta_i-\theta_j)+B\cos(\theta_i-\theta_j))\nonumber\\
&-B\|V_i\|^2
\end{align}

Taking the partial derivative of $Q_i$ with respect to $\|V_i\|$, we have the following:
\begin{align}
\frac{\partial Q_i}{\partial \|V_i\|}=\frac{Q_i-B\|V_i\|^2}{\|V_i\|}.
\end{align}
Since the loads and distribution lines in MGs are usually inductive rather than capacitive, $Q_i$ is usually larger than zero if CIGs share the reactive power properly. As $Q_i\geq0$, $B<0$ and $\|V_i\|^2\geq0$, we conclude that $Q_i$ has a positive correlation with $\|V_i\|$, i.e., $\|\boldsymbol{v_{o_i}^{dq}}\|$. Even if $Q_i<0$, the term $-B\|V_i\|^2$ may still make $Q_i-B\|V_i\|^2$ be positive. Note that $B$ is the susceptance of the output connector. By choosing $B$ properly, $Q_i-B\|V_i\|^2>0$ generally holds and thus $Q_i$ is still positively correlated with $\|\boldsymbol{v_{o_i}^{dq}}\|$.

Given the positive correlation between $CIG_i$ power injection and $\|\boldsymbol{v_{o_i}^{dq}}\|$, individual $\|\boldsymbol{v_{o_i}^{dq}}\|$ is regulated by controlling $P_i^{Vinv}$, which is one of the component of $P_i^*$, and $Q_i^{Vind}$, which is one of the component of $Q_i^{base}$, as follows:
\begin{align}
P_i^{Vind}&=K_p^{P}(V_s-\|\boldsymbol{v_{o_i}^{dq}}\|)+K_i^{P}\int_0^t(V_s-\|\boldsymbol{v_{o_i}^{dq}}\|)dt \label{P controller}\\
Q_i^{Vind}&=K_p^{Q}(V_s-\|\boldsymbol{v_{o_i}^{dq}}\|)+K_i^{Q}\int_0^t(V_s-\|\boldsymbol{v_{o_i}^{dq}}\|)dt. \label{Q controller}
\end{align}
The proportional gains and integral gains determine the extent of sacrificing the accuracy in active and reactive power sharing in the trade-off with individual $\|\boldsymbol{v_{o_i}^{dq}}\|$ regulation. $P_i^{Vind}$ and $Q_i^{Vind}$ are the stems of power sharing errors. The following is the two modes of operation.
\subsubsection{Partial Individual $\|\boldsymbol{v_{o_i}^{dq}}\|$ Control}
Integral gains are set to zero with at least one proportional gain being greater than zero. Trade-off between the accuracy in power sharing and individual $\|\boldsymbol{v_{o_i}^{dq}}\|$ regulation is determined by the value of $K_p^P$ and $K_q^Q$. Individual $\|\boldsymbol{v_{o_i}^{dq}}\|$ is not driven to $V_s$ due to the absence of the integral action.
\subsubsection{Exact Individual $\|\boldsymbol{v_{o_i}^{dq}}\|$ Control}
At least one integral gain is greater than zero. Individual $\|\boldsymbol{v_{o_i}^{dq}}\|$ is driven to $V_s$ due to the integral action. If $K_i^P$ and $K_p^P$ are set to zero and $K_i^Q$ is larger than zero, the performance of the controller will be very similar to those in \cite{I1,I2,L}, i.e., the active power is accurately shared and the accuracy in reactive power sharing is traded off with the exact individual $\|\boldsymbol{v_{o_i}^{dq}}\|$ regulation.

\begin{remark}
$Q-f$ instead of $P-f$ is used since we would like to control active power, $P$, as a non-slack variable for better coordination with primary energy source. Existing methods usually control $P$ as a slack variable and assume the primary energy source can supply active power instantaneously to regulate dc link voltage, which is not necessarily true. In the proposed control method, the output voltage magnitude and reactive power are slack variables. The amount of $P$ injection is directly controlled.
\end{remark}

\begin{remark}
Since distributed lines and loads are modelled as RL circuits, their active power consumption are voltage sensitive (their voltage depends on the active power supplied to them). As $P$ is a non-slack variable while reactive power, $Q$, is a slack variable, the most direct way to regulate average voltage is to control the $P$ setpoint of each CIG according to the active power sharing ratio. When the $P$ injection from a CIG changed, its $Q$ injection changes accordingly and may violate the $Q$ sharing ratio among CIGs. However, the error in $Q$ sharing is then elminated by the $Q-f$ droop control.
\end{remark}

\section{Case Study}
\begin{figure}[htbp]
\centerline{\includegraphics[scale=0.75]{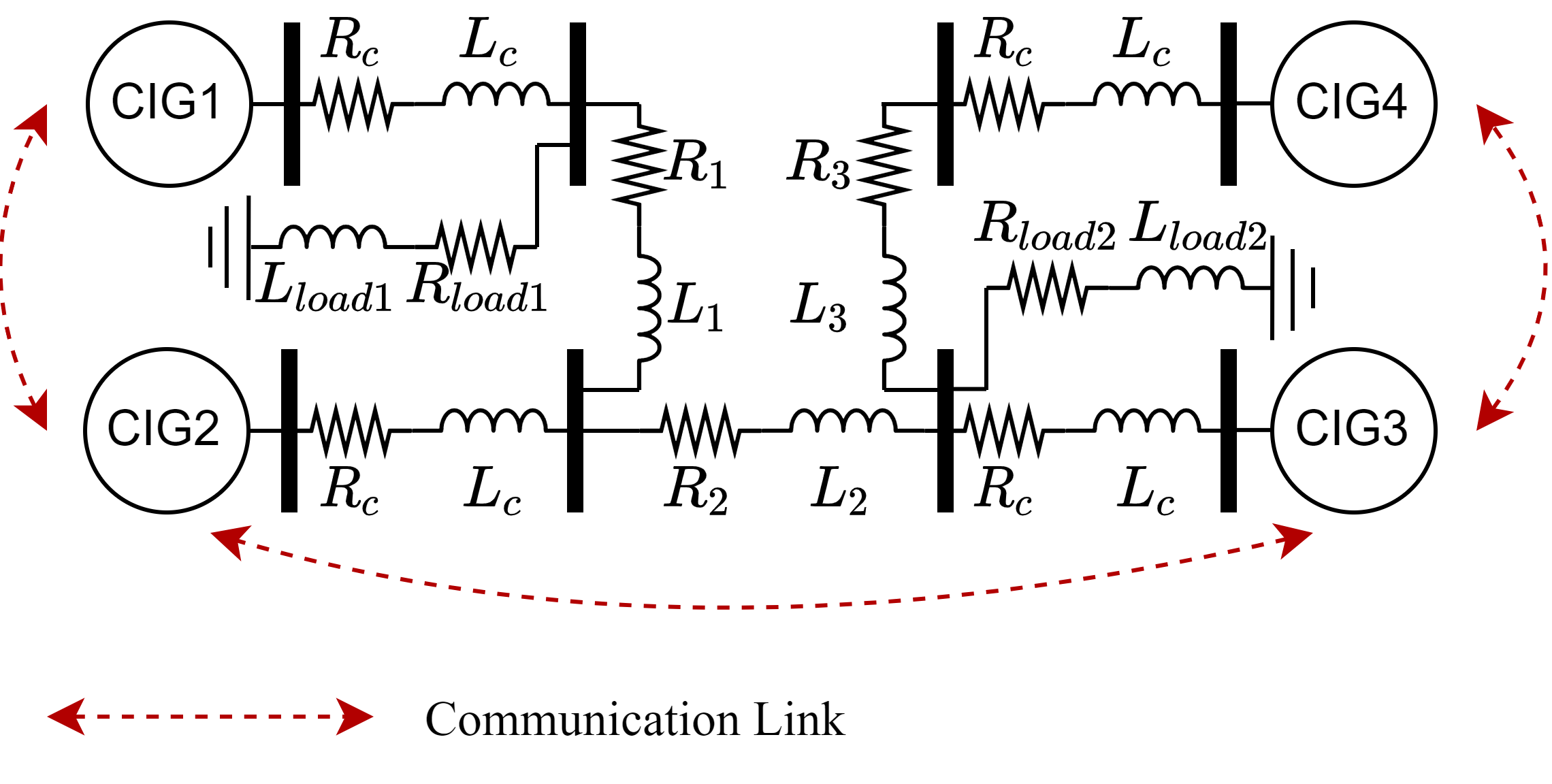}}
\caption{Microgrid Test System.}
\label{fig}
\end{figure}
\begin{figure*}[t]
\centering
\includegraphics[width=\linewidth]{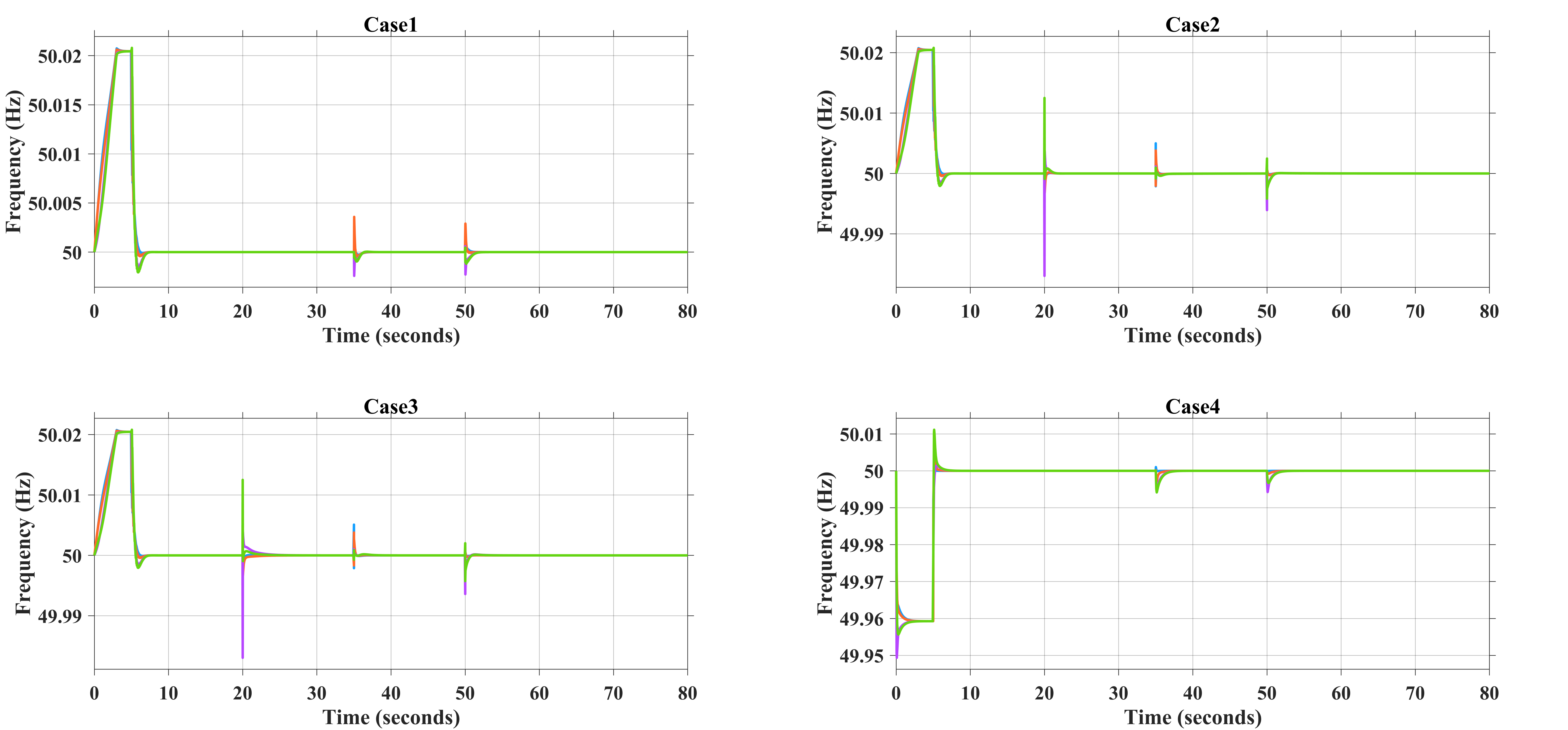}
\caption{Frequency. Blue: CIG1. Orange: CIG2. Purple: CIG3. Green: CIG4.}
\label{f}
\centering
\includegraphics[width=\linewidth]{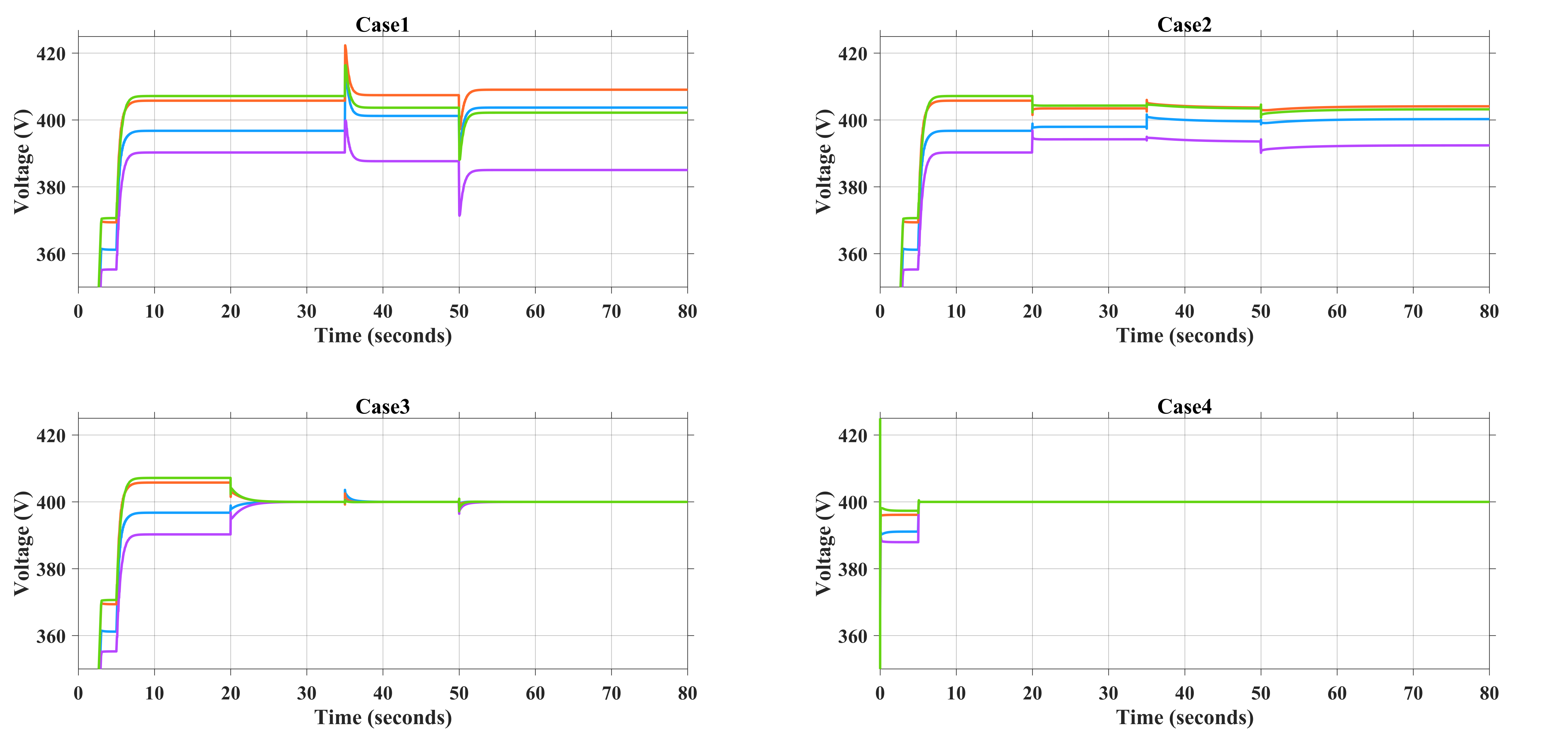}
\caption{Output voltage magnitude. Blue: CIG1. Orange: CIG2. Purple: CIG3. Green: CIG4.}
\label{v}
\end{figure*}

\begin{figure*}[t]
\centering
\includegraphics[width=\linewidth]{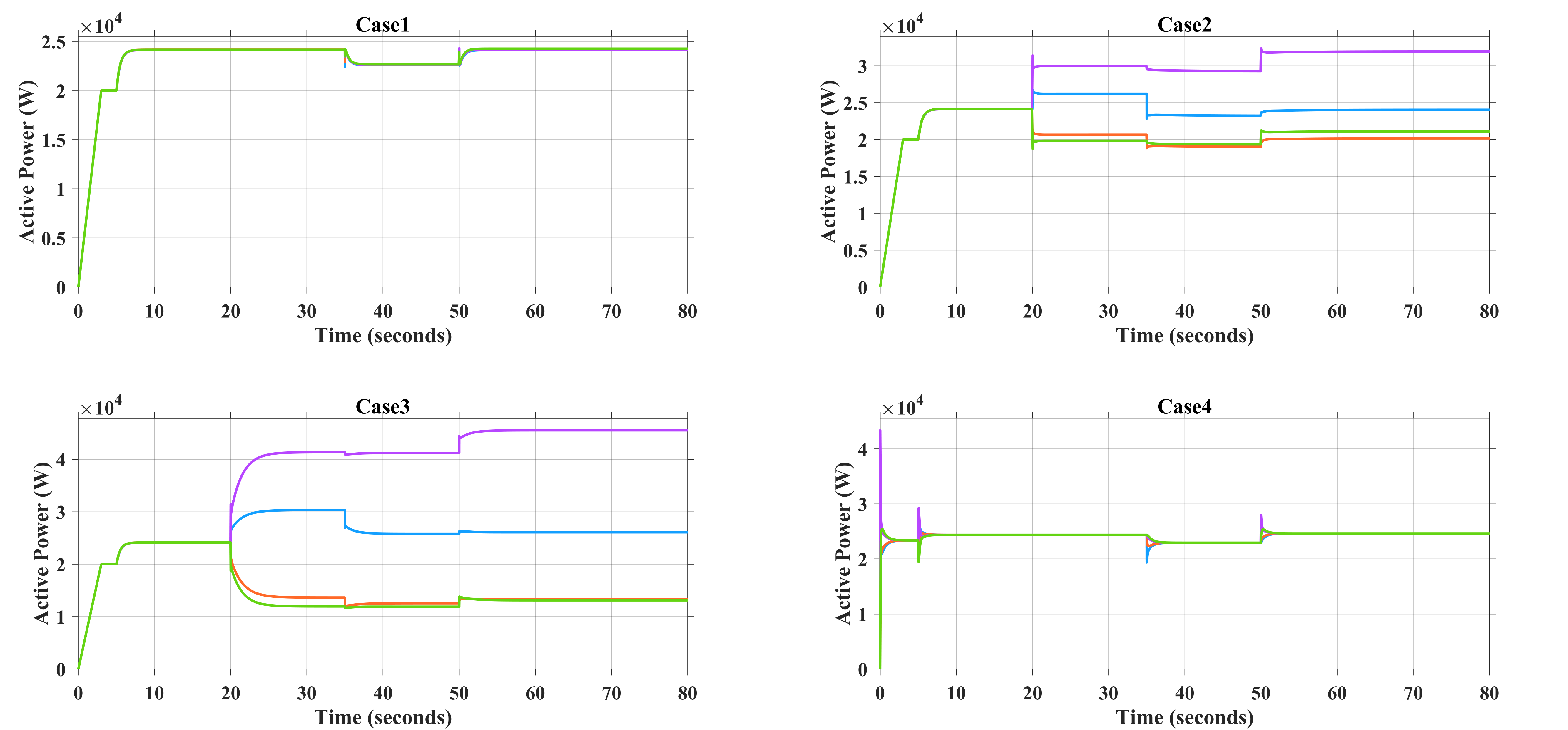}
\caption{Active power injection. Blue: CIG1. Orange: CIG2. Purple: CIG3. Green: CIG4.}
\label{p}
\end{figure*}

\begin{figure*}[t]
\includegraphics[width=\linewidth]{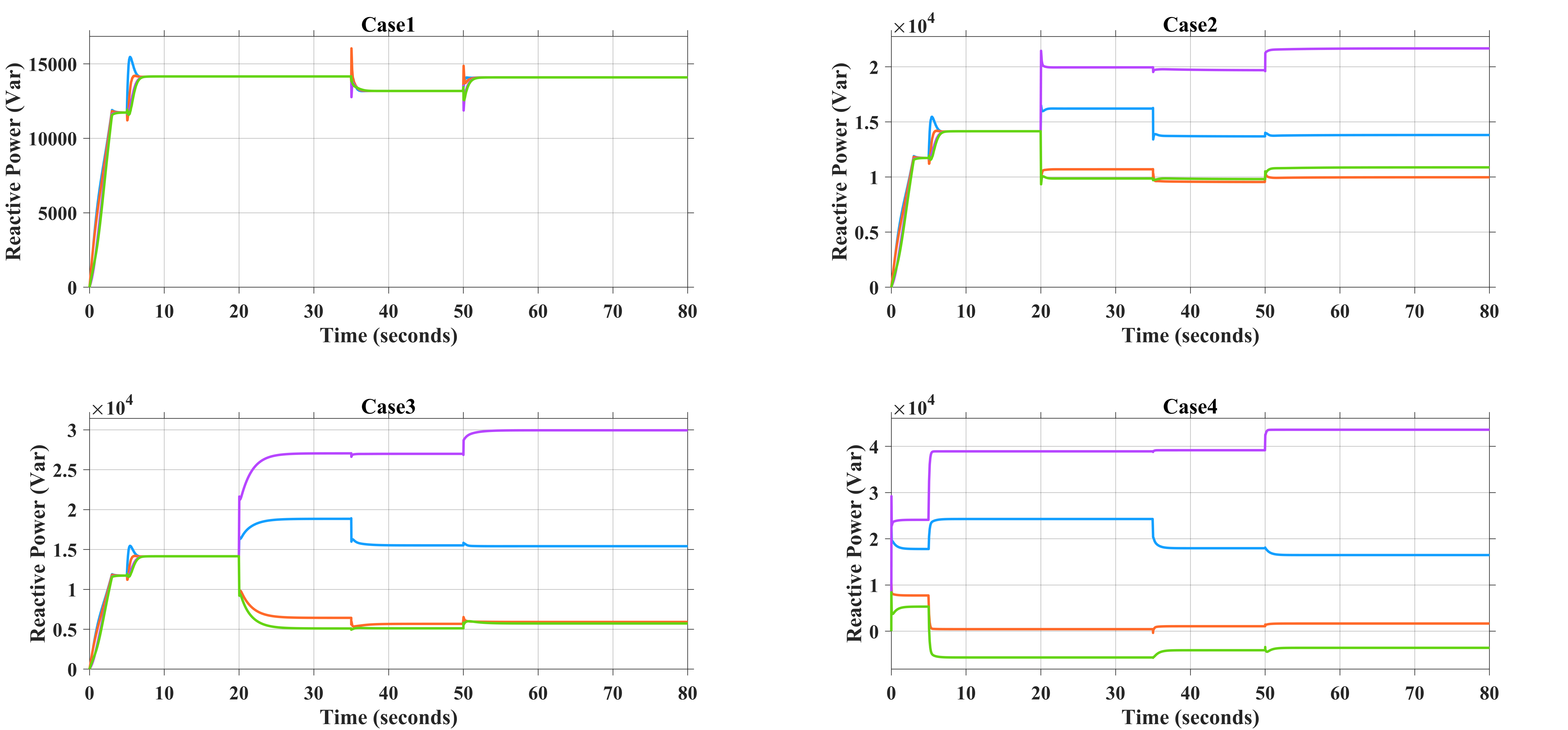}
\caption{Reactive power injection. Blue: CIG1. Orange: CIG2. Purple: CIG3. Green: CIG4.}
\label{q}
\end{figure*}


A test system originally from \cite{L} is modified for case study in this section. The digram of the test system is shown in Fig. 3. Four cases are simulated to compare the performance of the proposed control methods with the method used in\cite{I2,L} with the $P-f$, $Q-V$ droop control and secondary control (PFQV). Three different cases of the proposed control methods are used from case 1 to case 3, respectively. PFQV is implemented in case 4. Note that we assume continuous communication between CIGs in PFQV. This is different from case 1 to 3 where only discrete communication is used to demonstrate the robustness of the proposed controller. Besides, only one CIG in PFQV can access $V_s$ in PFQV but in the proposed control method we assume all CIGs know $V_s$. This assumption is valid as $V_s$ is usually defined in grid code and seldom changes. 

From case 1 to case 3, $Q_i^{nom}$ is set to 0Var and $P_i^{nom}$ is linearly increased from 0W to 20000W in first 3s in all CIGs. The power sharing gains $m_i$ and $n_i$ in all CIGs in each method are set to be the same to share the power equally among CIGs, i.e., the active and reactive power from each CIG should be the same ideally. In case 4, the droop gains in $P-f$ and $Q-V$ droop in all CIGs are set to be identical with each other as well. Therefore, their power sharing performance can be evaluated by observing whether the active and reactive power output from each CIG are the same. $V_s$ is chosen as 400V and two step load change occur at 35s in load1 and 50s in load2, respectively, in all cases. $t_1$ and $t_2$ are chosen as 0.025s and 0.05s, respectively.

From case 1 to case 3, the PI controller of $P_i^{Vav}$ is activated at 5s and the dynamic consensus algorithm is activated at 0s. For the decentralized individual CIG output voltage control, it is not triggered in case 1; mode 1 $(K_p^{P}=K_p^{Q}=1000)$ and mode 2 $(K_p^{P}=K_p^{Q}=1000$, $K_i^{P}=2000$, $K_i^{Q}=1500)$ are triggered in all CIGs at 20s in case 2 and case 3, respectively. Therefore, the control actions from case 1 to 3 are the same before 20s and the results overlap with each other. In case 4, both secondary frequency and voltage control are activated at 5s.

\begin{table}[hbtp]
\footnotesize
\centering
\settowidth\tymin{\textbf{Activities}}
\caption{Comparison of tradeoff of between different objectives in four cases}
\begin{tabulary}{\linewidth}{|C|C|C|C|}
\hline
&&&\\[-1.5em]
Case&Sacrfice of $P$ Sharing Accuracy&Sacrifice of $Q$ Sharing Accuracy& Voltage Regulation Extent\\
\hline
&&&\\[-1.5em]
1&No&No& Average Voltage Control\\
\hline
&&&\\[-1.5em]
2&Yes, via proportional control in \eqref{P controller}&Yes, via proportional control in \eqref{Q controller}& Average Voltage Control with Partial Individual $\|\boldsymbol{v_{o_i}^{dq}}\|$ Control\\
\hline
&&&\\[-1.5em]
3&Yes, via proportional and integral control in \eqref{P controller}&Yes, via proportional and integral control in \eqref{Q controller}& Average Voltage Control with Exact Individual $\|\boldsymbol{v_{o_i}^{dq}}\|$ Control\\
\hline
&&&\\[-1.5em]
4&No&Yes& Exact Voltage Regulation\\
\hline
\end{tabulary}
\end{table}

\subsection{Frequency}
Fig. \ref{f} compares the frequency regulation performance in four cases.  Their frequency regulation performance is similar with each other as the principle of secondary frequency control used is the same. The differences are that $Q-f$ primary droop control is used in case 1 to 3 while $P-f$ primary droop control is used in case 4 and the base value in $Q-f$ primacy droop control in case 2 and 3 started to change at 20s due to the decentralized individual CIG output voltage control, resulting in large disturbance in freuquency at 20s.

\subsection{Voltage Regulation}
Fig. \ref{v} compares the output voltage magnitude of CIGs in four cases. In case 1, only the average of output voltage magnitude is regulated to 400V. Individual CIG output voltage magnitude is not regulated. In case 2, individual CIG output voltage magnitude is partially regulated as we can seen each CIG output voltage magnitude is closer to 400V than that in Case 1 after enabling the partial individual $\|\boldsymbol{v_{o_i}^{dq}}\|$ control at 20s. In case 3, individual CIG output voltage magnitude is exactly regulated to 400V after enabling the exact individual $\|\boldsymbol{v_{o_i}^{dq}}\|$ control at 20s. In case 4, individual CIG output voltage magnitude is exactly regulated to 400V after the secondary voltage control in \cite{I2,L} is enabled at 5s. Noted that the voltage in case 4 is regulated more tightly than that in case 3 under step load changes. By selecting larger proportional gain and integral gain in decentralized individual CIG output voltage control, the voltage can be regulated more tightly in our proposed method. However, this may deteriorate the frequency regulation performance as the base value in the primary droop control is changed faster.

\subsection{Active Power}
Fig. \ref{p} compares the active power injection of CIGs in four cases. The order of active power sharing accuracy is : Case 1 $=$ Case 3 $>$ Case 2 $>$ Case 3.

\subsection{Reactive Active Power}
Fig. \ref{q} compares the reactive power injection of CIGs in four cases. The order of active power sharing accuracy is : Case 1 $>$ Case 2 $>$ Case 3 $>$ Case 4.

\subsection{Summary}

\begin{table}[hbtp]
\footnotesize
\settowidth\tymin{\textbf{Activities}}
\caption{Comparison of Relative Control Performance in different parameters in four cases}
\begin{tabulary}{\linewidth}{|C|C|C|C|C|}
\hline
&&&&\\[-1.5em]
Case&Frequency Regulation&$P$ Sharing Accuracy&$Q$ Sharing Accuracy& Exactness of Voltage Regulation\\
\hline
&&&&\\[-1.5em]
1&same&\checkmark\checkmark\checkmark&\checkmark\checkmark\checkmark\checkmark& \checkmark\\
\hline
&&&&\\[-1.5em]
2&same&\checkmark\checkmark&\checkmark\checkmark\checkmark& \checkmark\checkmark\\
\hline
&&&&\\[-1.5em]
3&same&\checkmark&\checkmark\checkmark&\checkmark\checkmark\checkmark\\
\hline
&&&&\\[-1.5em]
4&same&\checkmark\checkmark\checkmark&\checkmark& \checkmark\checkmark\checkmark\\
\hline
\end{tabulary}
\\[0.5em]
\footnotemark{The number of \checkmark only represents the relative control performance \\of that case on that parameter. It does not represent the relative performance across different parameters.}
\label{com}
\end{table}

Table \ref{com} summarizes the tradeoff in control performance between active power sharing accuracy, reactive power sharing accuracy and exactness of voltage regulation in different cases. It can be observed in Fig. \ref{p} and Fig. \ref{q} that the nearest CIG with respect to the load change reacts most if the control method cannot share active or reactive power accurately, i.e., when load 1 changes at 35s, CIG 1 power output (the blue lines) changes most. When load 2 changes at 50s, CIG 3 power output (the purple lines) changes most.

In practice, voltage regulation performance without decentralized individual CIG output voltage control (case 1) usually satisfies the voltage regulation requirement. Given a set of predefined active and reactive power sharing ratio among CIGs, exact voltage regulation may not always be possible. If exact voltage regulation is needed, the accuracy of power sharing has to be sacrificed. Different from existing controllers where the only choice to be scarified is reactive power sharing accuracy, the proposed controller allows the possibility sacrifice of active power sharing accuracy such that a balance can be stroked. Only sacrificing reactive power accuracy may lead to overloading of a particular CIG and high circulating current among CIGs.

Voltage may be more sensitive to active power change than reactive power change due to the high R/X ratio in MG \cite{A,L,H}, i.e., a small active power change may have the same effect on voltage as a large reactive power change. Hence, active power should also be considered in voltage regulation.

\section{Conclusion}
In this paper, we have proposed a tunable distributed secondary voltage and frequency regulation controller with controllable compromise between power sharing accuracy and voltage regulation. Different modes of operation has been demonstrated in the test system. Our case studies has shown that: 1) the proposed new control paradigm can overcome the inaccuracy in power sharing in conventional voltage droop meanwhile achieving accurate $\|\bar{v}_o\|$ regulation (case 1); 2) trade-off between active power sharing accuracy, reactive power sharing accuracy and voltage regulation performance is achieved (case 2 and 3); 3) reactive power sharing accuracy can be improved by including active power sharing in the trade-off (case 3 and 4). Future works will be done on choosing optimal parameters in the proposed controller.

\bibliographystyle{IEEEtran}
\bibliography{IEEEabrv,references}
\end{document}